# Striations in electronegative capacitively coupled radio-frequency plasmas: analysis of the pattern formation and the effect of the driving frequency

Yong-Xin Liu[1*], Ihor Korolov[2], Edmund Schüngel[3,4], You-Nian Wang[1],

Zoltán Donkó[2], and Julian Schulze[3,5]

[1]*Key Laboratory of Materials Modification by Laser, Ion, and Electron Beams (Ministry of Education), School of Physics and Optoelectronic Technology, Dalian University of Technology, Dalian 116024, China*
[2]*Institute for Solid State Physics and Optics, Wigner Research Centre for Physics, Hungarian Academy of Sciences, H-1121 Budapest, Konkoly-Thege Miklós str. 29-33, Hungary*
[3]*Department of Physics, West Virginia University, Morgantown, West Virginia 26506-6315, USA*
[4]*Evatec AG, Switzerland*
[5]*Institute for Electrical Engineering, Ruhr-University Bochum, 44801 Bochum, Germany*
*\*Current address: Plasma Processing Laboratory, Department of Chemical and Biomolecular Engineering, University of Houston, Houston, TX 77204-4004, USA*

## Abstract

Self-organized striated structures of the plasma emission have recently been observed in capacitive radio-frequency $CF_4$ plasmas by Phase Resolved Optical Emission Spectroscopy (PROES) and their formation was analyzed and understood by Particle in Cell / Monte Carlo Collision (PIC/MCC) simulations [Y.-X. Liu, et al. Phys. Rev. Lett. 116, 255002 (2016)]. The striations were found to result from the periodic generation of double layers due to the modulation of the densities of positive and negative ions responding to the externally applied RF potential. In this work, an in-depth analysis of the formation of striations is given, as well as the effect of the driving frequency on the plasma parameters, such as the spatially modulated charged species densities, the electric field, and the electron power absorption is studied by PROES measurements, PIC/MCC simulations, and an ion-ion plasma model. The measured spatio-temporal electronic excitation patterns at different driving frequencies show a high degree of consistency with the simulation results. The striation gap (i.e., the distance between two ion density maxima) is found to be inversely proportional to the driving frequency. In the presence of striations the minimum ($CF_3^+$, $F^-$) ion densities in the bulk region exhibit an approximately quadratic increase with the driving frequency. For these densities, the eigenfrequency of the ion-ion plasma is near the driving frequency, indicating that a resonance occurs between the positive and negative ions and the oscillating electric field inside the plasma bulk. The maximum ion densities in the plasma bulk are found not to exhibit a simple dependence on the driving frequency, since these ion densities are abnormally enhanced within a certain frequency range due to the ions being focused into the "striations" by the spatially modulated electric field inside the bulk region.

# 1. Introduction

The self-organized formation of spatial patterns of the light emission has been an attractive topic of fundamental research due to its remarkable influence on various plasma systems ranging from atmospheric to low pressure DC/RF discharges [1-10]. It is well known that the self-organized discharge structures can be generally divided into two broad categories. Multiple discharge channels, which are formed typically between two parallel plate electrodes and their relative locations are locked with respect to each other, belong to the first category. Such phenomena have been observed in laboratory plasmas over a wide pressure range, especially in atmospheric pressure dielectric barrier discharges [2-5]. Multilayer structures, appearing as alternating bright and dark areas along the length of a single plasma column correspond to the second category. They are usually termed as "striations", which have extensively been studied in atmospheric/low pressure and DC/RF glow discharges [6-9]. These structures are also expected to play an important role in a variety of plasma-based applications such as plasma enhanced chemical vapor deposition (PECVD) - often performed at relatively high pressures, low driving frequencies, and in electronegative gases - since the striations can drastically affect various process relevant plasma parameters, e.g., the flux-energy distribution functions of different charged species.

One of the most well-known scenarios is the self-organized striated structure occurring in the positive column region of dc glow discharges, wherein ion-acoustic or ionization waves are generally considered to be responsible for their appearance [1, 11-13]. Over the past few years, several types of striated structures have been observed in different versions of atmospheric pressure plasma jets (APPJ). Depending on the discharge conditions, the plasma streamer ejected from the nozzle was found in [14, 15] to be stratified into stationary bright and dark slender regions between the nozzle and the target plate. Very recently, a stratified filament occurring in the discharge tube of a RF argon APPJ was observed in numerical modeling, which identified the different nonlinear dependences of the ionization and recombination rates on the electron density as the reason for the formation of the spatial patterns [16]. In the low-pressure RF regime, stationary or moving striated structures at random intervals have also been observed in planar inductively coupled argon plasmas (ICP) [17-19]. The formation of these striations was inferred to arise from ionization waves similar to that in the positive column, also associated with the particular reactor geometry (e.g., an elongated tube or a narrow-gap ICP), but a comprehensive explanation is missing. Striations can also occur in plasma display panels [20], as well as in plasma clouds in the ionosphere [21, 22]. In these systems, the mechanisms of striation formation are still not fully understood, although efforts have been made to explain them via theories based on the electron kinetics. Several types of striations in electropositive RF discharges were analyzed by Penfold and coworkers, and they concluded that the underlying mechanisms in RF and DC glow discharges are similar [23].

Until recently, striated structures have never been observed in electronegative Capacitively Coupled RF (CCRF) plasmas, which have important applications in the semiconductor industry for etching and deposition processes [24]. In electropositive CCRF plasmas the RF voltage between the two electrodes drops primarily across the sheath regions, while the electric field strength in the bulk plasma is quite weak. The electrons gain energy primarily at the edges of the oscillating sheaths (α-mode) [25-28]. By increasing the working pressure and/or the applied voltage, the discharge can switch from the α-mode into the γ-mode, where secondary electrons emitted from the electrodes due to ion bombardment cause most of the ionization/excitation inside the sheath region. In these modes, the electron power absorption, as well as the ionization rates are typically high around the sheath edges and low in the plasma bulk [29, 30]. The electron power absorption mode in electronegative plasmas is more complex: electrons can also gain energy inside the bulk region [31-37], when the drift-ambipolar (DA) mode, analyzed in details in, e.g., [38-40], is effective. In this mode, the electrons gain energy from a drift field throughout the plasma bulk and from a steep ambipolar field at the edge of the collapsing sheath. It should be noted that in electronegative CCRF plasmas the drift field in the bulk region oscillates at the driving frequency. The (positive and negative) ions in this drift field are generally considered to be at rest (being non-responsive to the RF electric field alternating at, e.g., 13.56 MHz, due to their heavy masses), while the mobile electrons can instantaneously follow the rapidly alternating electric field and, thus, be accelerated to high energies and cause ionization/excitation. Consequently, in previous studies of such plasmas most of the attention has been paid to the investigation of the electron dynamics [41-45]. The visual appearance of such discharges typically shows a homogeneous bright zone in the central bulk region surrounded by two dark sheath regions adjacent to the electrodes.

At conditions when the ion plasma frequency becomes comparable to or higher than the driving frequency, positive and negative ions may respond to the RF electric field with an oscillating motion, generating space charges. This may lead to the formation of stable periodic structures ("striations") as explained below [8]. The electric field caused by the space charges enhances or attenuates the local drift electric field in the bulk, resulting in a spatially modulated electric field profile. The total field reinforces the response of the positive and negative ions to the alternating RF electric field by "focusing" them into the striations. Consequently, the space charges, as well as the striated electric field are enhanced due to a positive feedback. The effect is self-amplified until a periodic steady state is established, when a balance is achieved between two opposing processes: 1) the ion "focusing effect" and the ionization/attachment (leading to the formation of the positive and negative ions), and 2) the ion diffusion due to significant density gradients and the recombination of the positive and negative ions. The striated electric field results in the spatial modulation of the electron power absorption rate and, consequently, of the electron-impact excitation/ionization rate.

We have reported experimental observation of self-organized striated structures of the plasma emission for the first time in CCRF $CF_4$ plasmas driven at a frequency of 8 MHz by Phase Resolved Optical Emission Spectroscopy (PROES) [8]. These observations were reproduced by Particle in Cell /

Monte Carlo Collision (PIC/MCC) simulations. Moreover, an analytical model indicated that the resonance between the eigenfrequency of the ion-ion system and the driving frequency is the basis for the formation of the striated structures. Therefore, the value of the driving frequency is expected to play a crucial role for the properties of the striations. Its effects on various plasma parameters, e.g., the spatial profile of electric field, ion density, ionization/excitation patterns, etc., need to be explored.

In this work, we first investigate the basic properties of the striations, i.e., the spatially modulated electron power absorption and ionization/excitation dynamics, as well as the ion dynamics under the same conditions as in [8], but at a more detailed level, serving as the basis for an analysis of the effect of the driving frequency on various plasma parameters. Subsequently, we focus on the effects of the driving frequency on the plasma parameters, based on PROES measurements and PIC/MCC simulations. A model of the ion-ion plasma is also developed to aid explaining our results.

The paper is organized in the following way: in section 2 the experimental setup including diagnostics, the simulation method, and the ion-ion plasma model are described. In section 3, the results are discussed in two parts, as explained above. Finally, conclusions are drawn in section 4.

## 2. Experiment, simulation, and model
### 2.1 Experiment

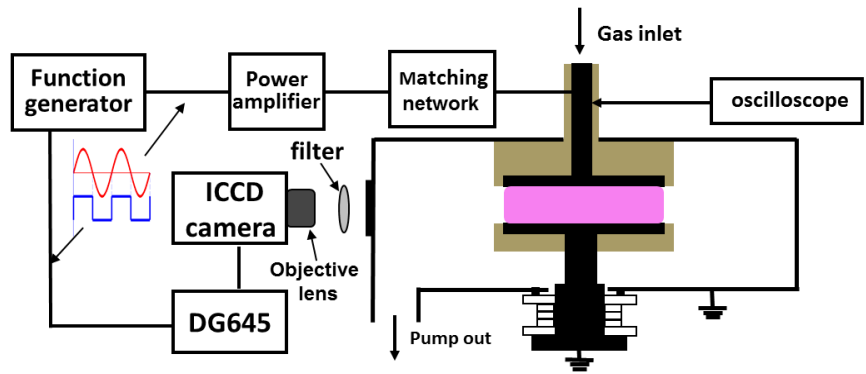

Figure 1. Scheme of the CCRF plasma chamber, supplemented with a phase resolved optical emission spectroscopy diagnostic system.

The scheme of the CCP reactor and the diagnostic systems used for the PROES measurements are shown in figure 1 (also described in [26]). The plasma is produced in $CF_4$ between two parallel circular electrodes made of stainless steel with a diameter of 10 cm. The gap between the electrodes is $L = 1.5$ cm and both electrodes are surrounded by Teflon. The chamber is made of stainless steel and is 28 cm in inner diameter. The bottom electrode and the chamber wall are grounded. A sinusoidal RF signal of a two-channel function generator (Tektronix AFG 3252C) is connected to a power amplifier (AR, Model 1000A225) and is applied to the top electrode via a matching network. Due to the geometric asymmetry of the reactor, the discharge is asymmetric as well, especially at low pressures, or high RF powers, leading to the generation of a significant DC self-bias voltage.

The function generator is also connected to a pulse delay generator (SRS INC., Model DG645) that triggers in a synchronized manner an intensified charge-coupled device (ICCD) camera (Andor iStar DH734) for the PROES measurements. The ICCD camera is equipped with an objective lens and an interference filter to detect the spatio-temporal emission intensity of a specifically chosen optical transition (at 585 nm) of Ne, which is admixed at 5% concentration as a tracer gas. The ICCD camera is fixed at about 1 m distance from the side quartz window of the chamber. From the light emission measurements, that are performed in a sequence through the RF period, the spatio-temporal electron-impact excitation rate from the ground state into the Ne2p$_1$-state is calculated based on a collisional-radiative model (for more details see [46, 47]).

The experiments are carried out for a frequency range of 4 MHz – 18 MHz, the gas pressure and the voltage amplitude are kept constant, respectively, at 100 Pa and 300 V. The voltage waveform is measured by a high-voltage probe (Tektronix P6015A) and is acquired with a digitizing oscilloscope (LeCroy Waverunner).

## 2.2 Simulations

Our numerical studies are based on a bounded 1D3V Particle in Cell simulation code, complemented with a Monte Carlo type treatment of collision processes (PIC/MCC), see e.g. [48]. The code considers one spatial (axial) coordinate and is three-dimensional in the velocity space. The electrodes of the discharge are assumed to be plane and parallel. The cross sections of electron-CF$_4$ collision processes are taken from [49], with the exception of the electron attachment processes (producing $CF_3^-$ and $F^-$ ions), for which we use data from [50]. We use an extensive set of electron-impact collision processes, however, disregard many of the products created in these reactions. The only products considered are $CF_3^+$, $CF_3^-$, and $F^-$ ions, which play the most important role in CF$_4$ discharge plasmas. These ions can participate in various ion-molecule reactive processes, as well as in elastic collisions [51-53]. The ion-ion recombination rate coefficients are taken from [54], while the rate for the electron- $CF_3^+$ recombination process is from [55]. In the simulations we assume a gas temperature of 350 K. We neglect the ion-induced emission of secondary electrons from the electrodes in order to simplify the analysis. We assume that electrons reaching the electrodes are reflected with a probability of 0.2 [56]. More details of our model, tables and graphical representations of the cross sections can be found in previous publications, e.g., [57]. The simulations make it possible to determine in a self-consistent manner the spatio-temporal distributions of several discharge characteristics, like the electric potential and field strength, particle densities, fluxes, velocities, as well as reaction (e.g. excitation, ionization, attachment, etc.) rates, which provide a deep insight into the physics of the plasma considered.

In the simulations we use 600 grid points to resolve properly the inter-electrode gap and between 15 000

and 30 000 time steps within an RF period, depending on the driving frequency, to resolve properly the temporal dynamics of all plasma species and to fulfill the relevant stability criteria of the numerical method.

The PIC/MCC simulations are carried out for parameter sets matching the experimental conditions. It should be kept in mind, however, that due to the asymmetry of the experimental electrode configuration, the measured voltage waveforms exhibit a large negative self-bias for some conditions, which cannot be accounted for by the simulations. In all cases the amplitudes of the measured voltage waveforms are used as input for the PIC/MCC simulations.

We also note that comparisons will be made between the spatio-temporal distributions of the excitation rate measured in the experiments and the ionization rate calculated in the simulations. This approach is justified by the fact that both of these processes represent the dynamic behavior of electron groups with similar energies, due to their comparable threshold energies (the excitation threshold of the $2p_1$ state of neon is ~19 eV, while the ionization threshold of $CF_4$ molecules is ~16 eV).

## 2.3 Ion-ion plasma model

To understand the underlying physics of the formation of striations, we analyze the coupling between the motion of the positive and negative ions in an ion-ion plasma under the influence of an RF electric field by solving analytically a one-dimensional uniform slab ion-ion plasma model in this section.

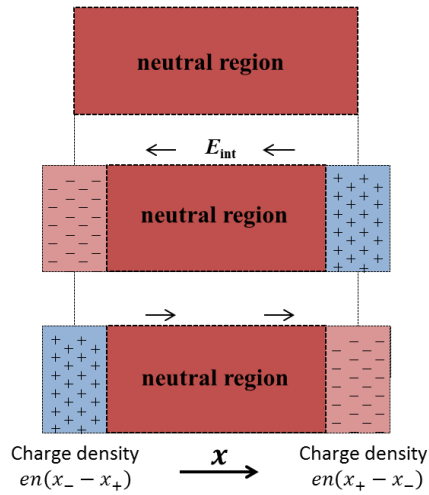

Figure 2 Schematic of the coupled motion of the positive and negative ion system (with equal positive and negative ion densities, $n_+ = n_-$) immersed in a time-varying, spatially uniform electric field $E_{ext} = E_0\cos(\omega_{rf}t)$, where $E_0$ is a constant. $x_\pm$ indicates the displacements of positive and negative ion groups, respectively. Note that the displacement of the ions moving towards the right is defined to be positive.

The motion of the positive and negative ions can be determined from the momentum balance equation,

$$mn[\frac{\partial u}{\partial t} + u\frac{\partial u}{\partial x}] = qnE - \frac{\partial p}{\partial x} - mn\nu u \qquad (1)$$

where $m$ is the ion mass, $n$ is the ion number density, $u$ is the ion mean velocity, $q$ is the ion charge, $E$ is the electric field, $p$ is the ion pressure, and $\nu$ is the ion collision frequency. For 100 Pa pressure, the term $mnu\frac{\partial u}{\partial x}$ is small compared to the collision term ($mn\nu u$), so we neglect this term on the left hand side of eq. (1). As we consider uniform positive and negative ion densities, the ion pressure gradient vanishes ($\frac{\partial p}{\partial x} = 0$), as well and eq. (1) reduces to

$$\dot{u} + \nu u = \frac{q}{m}E \qquad (2)$$

The electric field $E$ is composed of an "external" field component $E_{ext}$ and an "internal" field component $E_{int}$. If the positive ions (with density $n_+$) move a distance $x_+$ and the negative ions (with density $n_-$) move a distance $x_-$, the space charges of $en_+(x_+ - x_-)$ and $en_-(x_- - x_+)$, respectively, form at the edges of the slab, as shown in figure 2. Note that the displacement of the ions moving towards the right is defined to be positive. The space between the two regions with space charges is filled by quasi-neutral plasma, wherein the electric field is given as:

$$E = E_{ext} + E_{int} = E_0 \cos(\omega_{rf} t) - \frac{qn}{\varepsilon_0}(x_+ - x_-), \qquad (3)$$

where $\omega_{rf}$ is the angular frequency of the external electric field. We adopt equal positive and negative ion densities, i.e., $n_+ = n_-$, so that equations (2) and (3) yield

$$\ddot{x}_- + \nu \dot{x}_- - \omega_-^2 (x_+ - x_-) = -\frac{e}{m_-} E_0 \cos(\omega_{rf} t) \qquad (4)$$

for the negative ions and

$$\ddot{x}_+ + \nu \dot{x}_+ + \omega_+^2 (x_+ - x_-) = \frac{e}{m_+} E_0 \cos(\omega_{rf} t) \qquad (5)$$

for the positive ions.

These two coupled equations describe the motion of both ion species in a RF oscillating electric field. By subtracting equation (4) from (5), we obtain

$$(\ddot{x}_+ - \ddot{x}_-) + \nu(\dot{x}_+ - \dot{x}_-) + (\omega_+^2 + \omega_-^2)(x_+ - x_-) = \frac{m_+ + m_-}{m_+ m_-} eE_0 \cos(\omega_{rf} t). \qquad (6)$$

We define $x = x_+ - x_-$, $\omega^2 = \omega_+^2 + \omega_-^2 = \frac{e^2 n}{\varepsilon_0 m_+} + \frac{e^2 n}{\varepsilon_0 m_-} = \frac{e^2 n}{\varepsilon_0 \mu}$, $\beta = \frac{eE_0}{\mu}$, and $\mu = \frac{m_+ m_-}{m_+ + m_-}$, where $x$ is the relative displacement of the positive and negative ions, $\omega$ is the eigenfrequency of the positive and negative ion plasma, and $\mu$ is the reduced mass of the positive and negative ions.

The coupled motion of positive and negative ions can be considered as a system, so equation (6) can be reduced to:

$$\ddot{x} + \nu \dot{x} + \omega^2 x = \beta \cos(\omega_{rf} t). \qquad (7)$$

In order to obtain an analytical solution, we assume that the ion collision frequency, $\nu,$, is a constant.

Assuming an initial relative velocity $\dot{x}(0) = 0$ and an initial relative displacement $x(0) = 0$, the following analytical solution of equation (7) can be obtained:

$$x(t) = \frac{\beta}{\sqrt{\left(\omega_{rf}^2 - \omega^2\right)^2 + \omega_{rf}^2 \nu^2}} [\sin(\omega_{rf}t - \varphi) - \frac{\omega}{\theta} \cdot e^{-\frac{\nu}{2}t} \cdot \sin(\theta t - \vartheta)], \qquad (8)$$

with $\cos\varphi = \frac{\omega_{rf}\nu}{\sqrt{\left(\omega_{rf}^2 - \omega^2\right)^2 + \omega_{rf}^2 \nu^2}}$, $\cos\vartheta = \frac{\nu/\omega \cdot (\omega_{rf}^2 + \omega^2)}{\sqrt{\left(\omega_{rf}^2 - \omega^2\right)^2 + \omega_{rf}^2 \nu^2}}$, and $\theta = \frac{1}{2}\sqrt{4\omega^2 - \nu^2}$,

where θ is a real number, since we have $4\omega^2 - \nu^2 > 0$ for the conditions investigated here.

When the system reaches a stable state (i.e., t → ∞, and $e^{-\frac{\nu}{2}t} \to 0$), we have

$$x(t) = \frac{eE_0/\mu}{\sqrt{\left(\omega_{rf}^2 - \omega^2\right)^2 + \omega_{rf}^2 \nu^2}} \sin(\omega_{rf}t - \varphi) = A\sin(\omega_{rf}t - \varphi), \qquad (9)$$

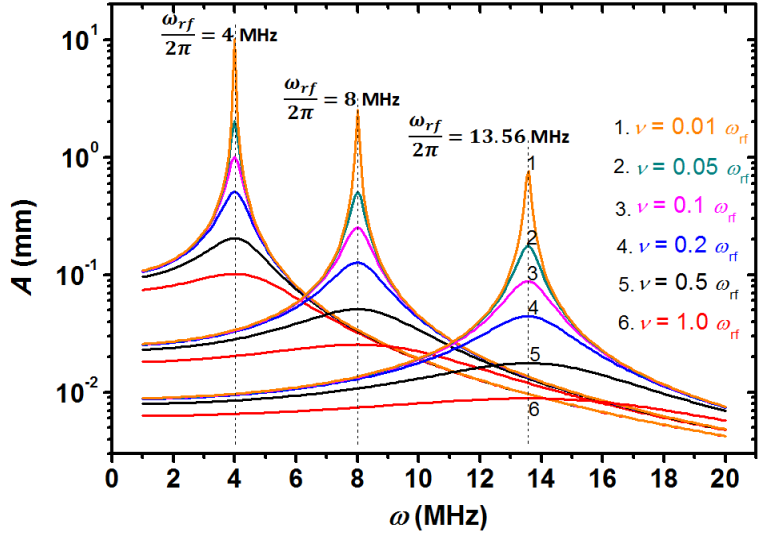

Figure 3. The amplitude $A$ appearing in equation (9) as a function of the eigenfrequency $\omega$ at different collision frequencies $\nu$ ($0.01\omega_{rf} \sim \omega_{rf}$). The results for three typical angular frequencies of the external electric field are compared, i.e., $\omega_{rf} = 2\pi \times 4 \times 10^6$ s$^{-1}$, $2\pi \times 8 \times 10^6$ s$^{-1}$, and $2\pi \times 13.56 \times 10^6$ s$^{-1}$, assuming $E_0 = 10^4$ V/m for the external electric field.

The amplitude $A$ in equation (9), as a function of the eigenfrequency $\omega$, at several values of the collision frequency, $\nu$, and three typical angular frequencies of the external electric field, $\omega_{rf}$, is shown in figure 3. At $\omega = \omega_{rf}$ the amplitude of the relative replacement of the ions exhibits a maximum, indicating that a resonance occurs between the ion-ion plasma and the external electric field. Under the resonance condition, the relative displacement reduces to

$$x(t) = \frac{eE_0/\mu}{\omega_{rf}\nu} \sin(\omega_{rf}t), \qquad (10),$$

having an amplitude $A^* = \frac{eE_0/\mu}{\omega_{rf}\nu}$. Both an increasing pressure and an increasing driving frequency $\omega_{rf}$

suppress the oscillation of ions ($A, A^* \to 0$, see figure 3). At the base case in this study ($\omega_{rf} = 2\pi \times 8 \times 10^6$ s$^{-1}$, $\nu = 0.5\omega_{rf} = 25.12 \times 10^6$ s$^{-1}$ (if we assume an ion velocity of $v = 1.1 \times 10^3$ m/s, $p = 100$ Pa,), the amplitude is $A = 0.051$ mm at the resonance.

Although the slab ion-ion plasma model does not predict the periodicity of the striations, it reveals the basic physical mechanism behind their formation. Most importantly, it predicts the resonance condition of the positive and negative ion plasma with the RF electric field, i.e., $\omega_{rf} = \omega_+ + \omega_- = \sqrt{\frac{e^2 n}{\varepsilon_0 \mu}}$. Based on this resonance condition we can obtain the critical ion density, $n_{\text{critical}} = \omega_{rf}^2 \varepsilon_0 \mu / e^2$, beyond which the striations can be present. Besides, the amplitude of the relative displacement of the positive and negative ions, $A^* = \frac{eE_0/\mu}{\omega_{rf}\nu}$, is correlated with the dependence of the striation gap on external control parameters, i.e., the driving angular frequency $\omega_{rf}$, the working gas pressure (via the collision frequency, $\nu$), the gas type (that defines the reduced ion mass), and the bulk electric field $E_0$. Both, the striation gap and the amplitude of the ion displacement are found to be inversely proportional to the driving frequency.

## 3. Results

This section consists of two parts: (i) the spatially modulated charged species densities, electric field, electron power absorption rate, and other plasma parameters at the base case (8 MHz, 300 V, and 100 Pa) are discussed in section 3.1. (ii) The effects of the driving frequency on the striated structures are studied in section 3.2.

### 3.1 Properties of the striations at the base case (8 MHz, 300 V, 100 Pa)

Figures 4 (a, b) show the spatio-temporal distributions of the measured electron-impact excitation rate and the simulated electron-impact ionization rate, respectively, under the conditions mentioned above. The total electron-impact excitation rate, the dissociative attachment rate (corresponding to the formation of F$^-$ ions), the net charge density, the electric field, the electron power absorption rate and the electron mean energy are presented in figures 4 (c ~ h). One can see from figures 4 (a, b) that the experimentally determined electron-impact excitation rate is high near the grounded electrode during the first half of the RF period and near the powered electrode during the second half. The intensive excitation is concentrated within a relatively small spatio-temporal region close to the sheath edge at its collapsing phase. Particularly, the measured excitation rate exhibits a remarkable spatially modulated structure, suggesting that the number of electrons with energies above the excitation threshold of 19 eV (Ne 2p$_1$-state) varies periodically in space. In the experiment these striations are stable and can directly be observed by eye. These experimentally observed patterns agree qualitatively well with the layered spatio-temporal distribution of the ionization rate obtained from the simulations (figure 4(b)). This allows us to further explore the details of the striations in electronegative CCRF $CF_4$ discharges by PIC/MCC simulations.

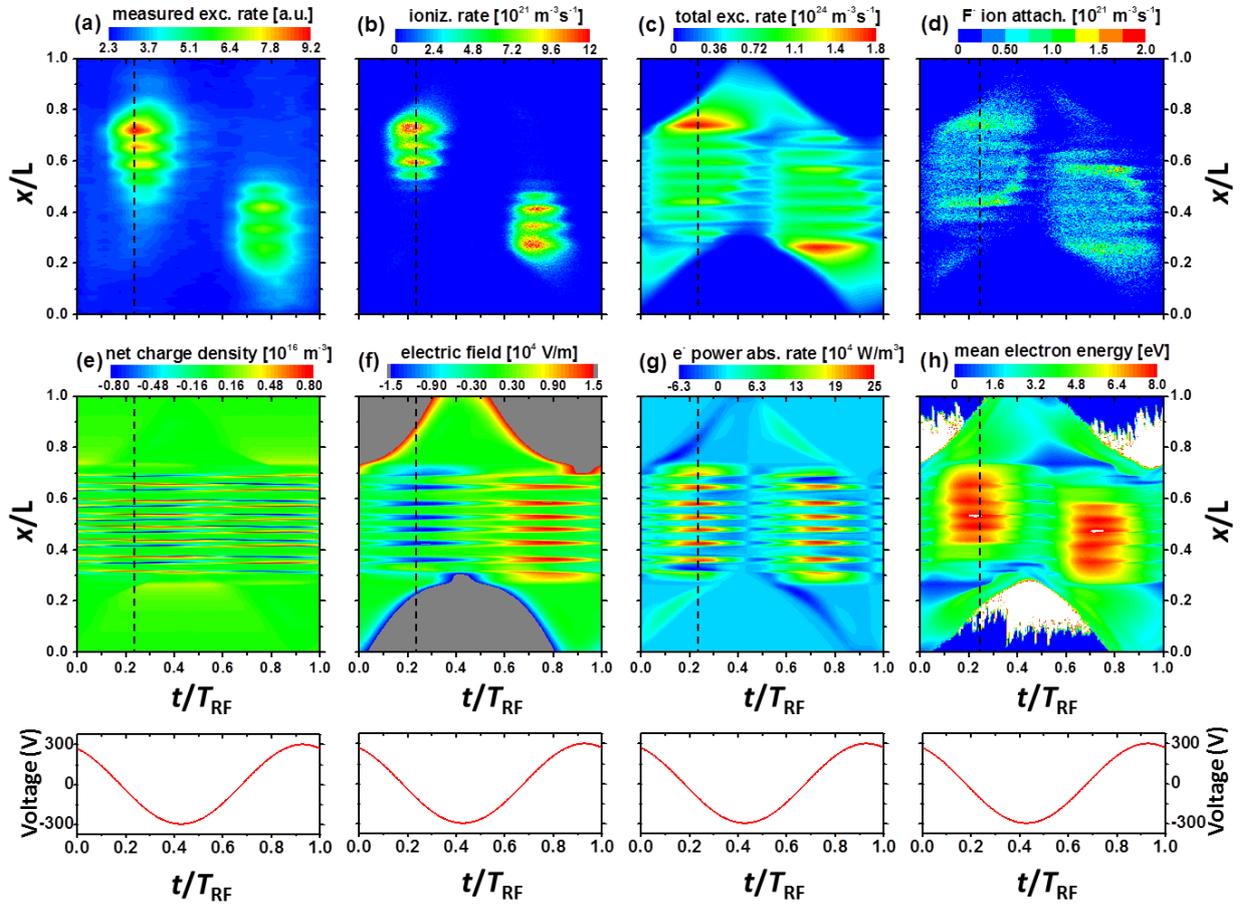

Figure 4. PROES results: spatio-temporal plot of the electron-impact excitation rate (a). PIC/MCC simulation results: spatio-temporal plots of the electron-impact ionization rate (b), excitation rate (c), dissociative attachment rate of $F^-$ ion formation (d), net charge density (e), electric field (f), electron power absorption rate (g), mean electron energy (h). The results are presented for 8 MHz, 300 V, and 100 Pa. The powered electrode is at $x/L = 0$, while the grounded electrode is at $x/L = 1$, $T_{RF}$ indicates one RF period. The bottom panels show the driving voltage waveform and the vertical dashed lines indicate a characteristic time within the RF period, at which plasma parameters will be analyzed in more detail in the frame of fig. 5.

The spatially modulated excitation/ionization rate is caused by electrons accelerated by the spatially modulated electric field profile (figure 4(f)), which leads to a spatially modulated electron power absorption rate (figure 4(g)). It can be seen in figure 4(f) that the electric field is quite different from the previously observed homogeneous drift field inside the bulk region of an electronegative CCRF plasma [27-29], as it is a superposition of the drift field generated by the externally applied voltage and the electric field caused by the space charge (figure 4(e)). The spatially alternating space charge has been found to be a consequence of the periodic motion of the positive and negative ions into opposite directions due to their response to the RF modulated local electric field [8]. To be more specific, during the first half of one RF period, a negative electric field inside the bulk region accelerates the positive ions towards the bottom (powered) electrode and the negatively charged ions towards the top (grounded) electrode. Thus, space charge builds up locally with alternating sign at the positions of highest ion density gradient. This can clearly be seen in figure 5(a), which presents the axial profiles of the charged species ($CF_3^+$, $F^-$, $CF_3^-$ ions, electron) densities, the net charge density and the electric field at a certain time within one RF period indicated by the vertical dashed lines in figure 4. These space charges, in turn, generate an electric field, which is comparable in magnitude with the drift electric field and enhances or attenuates the local field. The resulting total electric field is dominated by a striated pattern (see figures 4(f) and 5(a)). Note that under the conditions of the base case the local electronegativity, $\xi = n_-/n_e$, is within the range, $38 < \xi < 205$, with the minimum electronegativity $\xi_{min}$ occuring at the positions of the minimum ion density and the maximum electronegativity $\xi_{max}$ at the positions of the maximum ion density ($n_-$ and $n_e$ are the local negative ion density and the local electron density, respectively). Thus, the electron density (see figure 5(a)) in the plasma bulk is so low that it hardly contributes to the space charge distribution [26,38,39]. The ion motion in response to the externally applied potential variations is hardly affected by the electrons. Therefore, the ion-ion plasma model introduced in section 2.3 can be used to describe the ion motion, which leads to the formation of the space charges and is, thus, the origin of the formation of the striations. However, despite their low density the electrons play a crucial role for the sustainment of the striations, since electron impact ionization of the neutral gas compensates the charged particle losses.

This spatially modulated structure of the excitation/ionization rate can also be observed in the spatio-temporal distributions of the total electron excitation and dissociative attachment rates obtained from the simulations (figures 4(c, d)). However, the difference is that the spatio-temporal distributions of the electron-impact excitation and dissociative attachment rates spread all over the entire bulk region for each half of one RF period, primarily due to the lower energy thresholds of these reactions. To be specific, the electron impact ionization threshold of $CF_4$ is relatively high (~16 eV), and it takes a longer distance for the electrons to get accelerated to the threshold energy by the bulk electric field, therefore intensive ionization starts only at $x/L = 0.5$ and continues up to the position of the collapsing sheath edge. In contrast, the total electron-impact excitation and attachment reaction thresholds (7.5 eV and 5 eV) are relatively low, and the portion of electrons with energies above these thresholds is much higher, so that the electron-impact excitation and attachment reactions can start at the location closer to the expanding sheath edge (see figures 4(c, d)). Besides, by comparing the results in figures 4(b, c, and d), we find that the maxima of the electron-impact ionization, excitation and dissociative attachment rates have different orders of magnitude, due to different thresholds, as well as due to different cross sections for the different reaction types. In particular, the maximum of the total excitation rate is more than two orders of magnitude higher than the maximum of the ionization rate, which is about 6 times higher than that of the dissociative attachment rate.

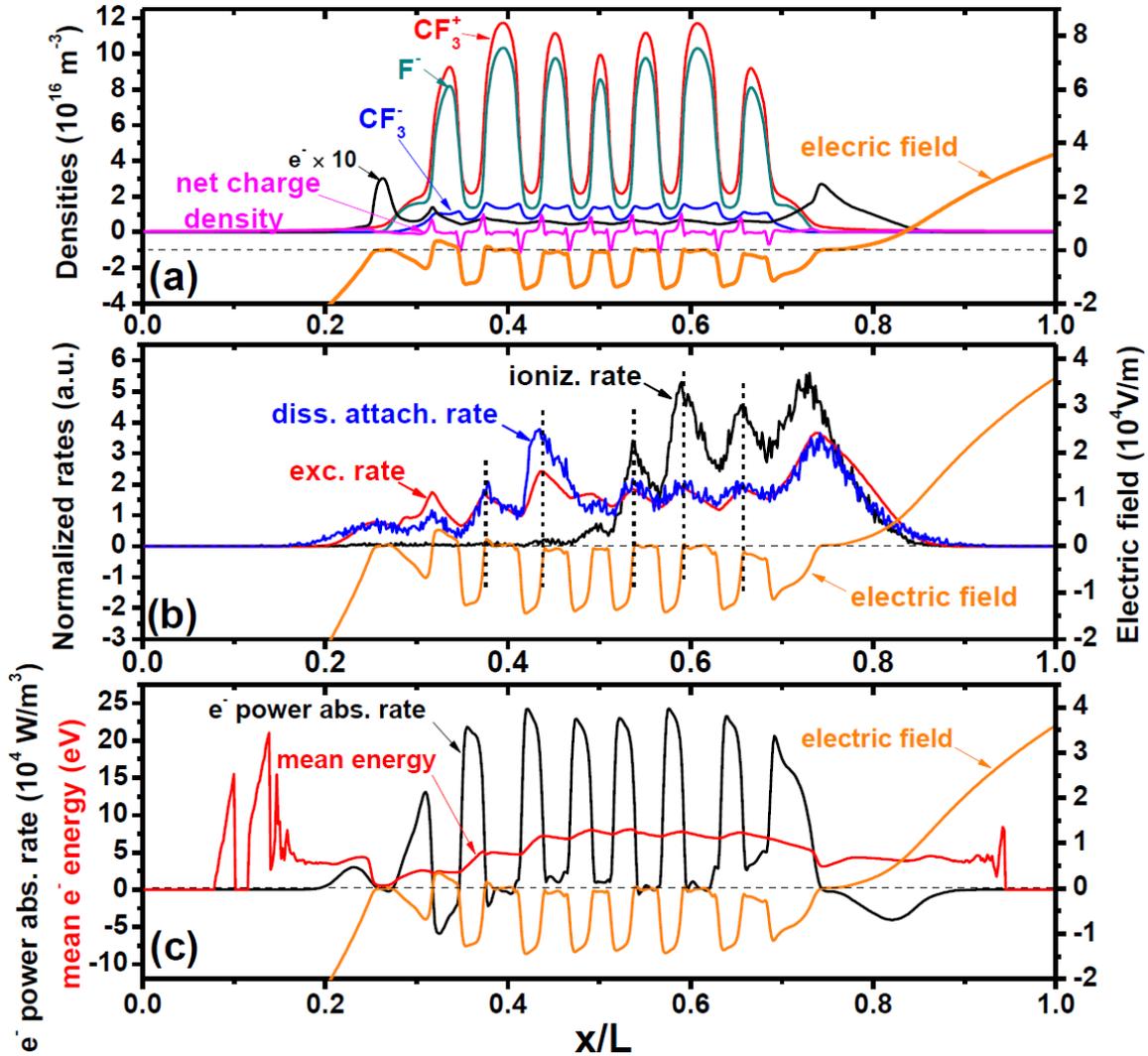

Figure 5 PIC/MCC simulation results: axial profiles of charged species densities and net charge density (a) at the time within one RF period corresponding to the vertical dashed lines in figures 4 and 6 (a1, b1, and c1). Axial profiles of electron-impact ionization, excitation, and dissociative attachment ($F^-$ formation) rates (b), electron power absorption rate and mean energy (c) at the time within one RF period corresponding to the vertical dashed lines in figures 4 and 6 (a1,b1, and c1). Note that the electric field profile at the same time within one RF period is inserted in each figure. The electron density is multiplied by a factor of 10. The axial profiles of the excitation, ionization and dissociative attachment rates are normalized by their space-averaged values. The discharge conditions are the same as in figure 4.

Although electrons hardly affect the net charge density distribution, they play a key role in the formation and the sustainment of the striated structure in the spatio-temporal distributions of the ionization, excitation, and dissociative attachment rates. In order to illustrate this effect the profiles of various plasma parameters are shown in figure 5 at the time indicated by the vertical dashed lines in figure 4. As it can be seen from figure 5(b) the maxima of the ionization, excitation and dissociative attachment rates always occur at the same position, i.e., at the right edges of the narrow regions of strong negative electric field (indicated by the vertical dashed lines in figure 5 (b)). At the first half of one RF period, the electrons are accelerated towards the grounded electrode by the bulk electric field, and the electron power absorption rate is strongly modulated in space, as shown in figure 5(c). It is clear that the electron power absorption rate ($P_e = J_e \cdot E$, where $J_e$ and $E$ are the electron current density and the electric field, respectively) is high when $|E|$ is high inside the bulk region, due to the almost uniform axial distribution of the electron flux $\Gamma_e$ inside the bulk region (see figure 6(c2)). During the time of high bulk electric field within one RF period an electron crosses a distance of multiple striation gaps. It accumulates energy, while it moves through regions of high electric field in between the striations, and loses energy, while it moves through regions of vanishing electric field at the positions of the density maxima. Overall, the mean electron energy inside the striation gaps increases as a function of position until significant ionization/excitation is caused (see figure 5). This mechanism leads to an almost constant mean electron energy with small modulations in space. A more detailed analysis shows that the maximum electron energy gain within one region of strong electric field in between two striations of width $\Delta x \approx 0.44$ mm is $\varepsilon_{max} = eE\Delta x \approx 6$ eV. At 100 Pa $\Delta x$ is comparable to the electron mean free path (for an electron energy between 10 and 20 eV).

Figures 6 (a1 and b1) present the spatio-temporal distributions of the $CF_3^+$ ion and $F^-$ ion densities for the base case. The positive ($CF_3^+$) and negative ($F^-$) ions move such a small distance per half rf period due to their large masses, that one cannot observe the ion motion in the spatio-temporal ion density distributions. However, the periodic motions of the positive and negative ions into opposite directions can clearly be seen in figures 6 (a2 and b2), which depict the $CF_3^+$ and $F^-$ ion fluxes. It is clear that as a consequence of their reaction to the local RF electric field inside the bulk plasma, the positive and negative ions move in opposite directions and the ion fluxes lag behind the electric field by a phase of π/2 due to the ion inertia. We see from figures 6 (a3) and (b3) that the $CF_3^+$ and $F^-$ ion mean velocities, whose profiles look quite different from the ion flux profiles, are approximately proportional to the local electric field strength, because the ions move a short distance per half rf period (the maximum displacement of a $F^-$ ion per half RF period is $x_{max,F^-} \approx 0.065$ mm, which is significantly smaller than the striation gap) so that they can only experience the local electric field. Note that the maximum displacement of a $F^-$ ion is comparable to the model result of $x_{model} \approx 0.051$ mm, which is calculated based on the ion-ion plasma model (section 2.3) with $\omega_{rf} = 2\pi \times 8 \times 10^6$ s$^{-1}$ and $p = 100$ Pa.

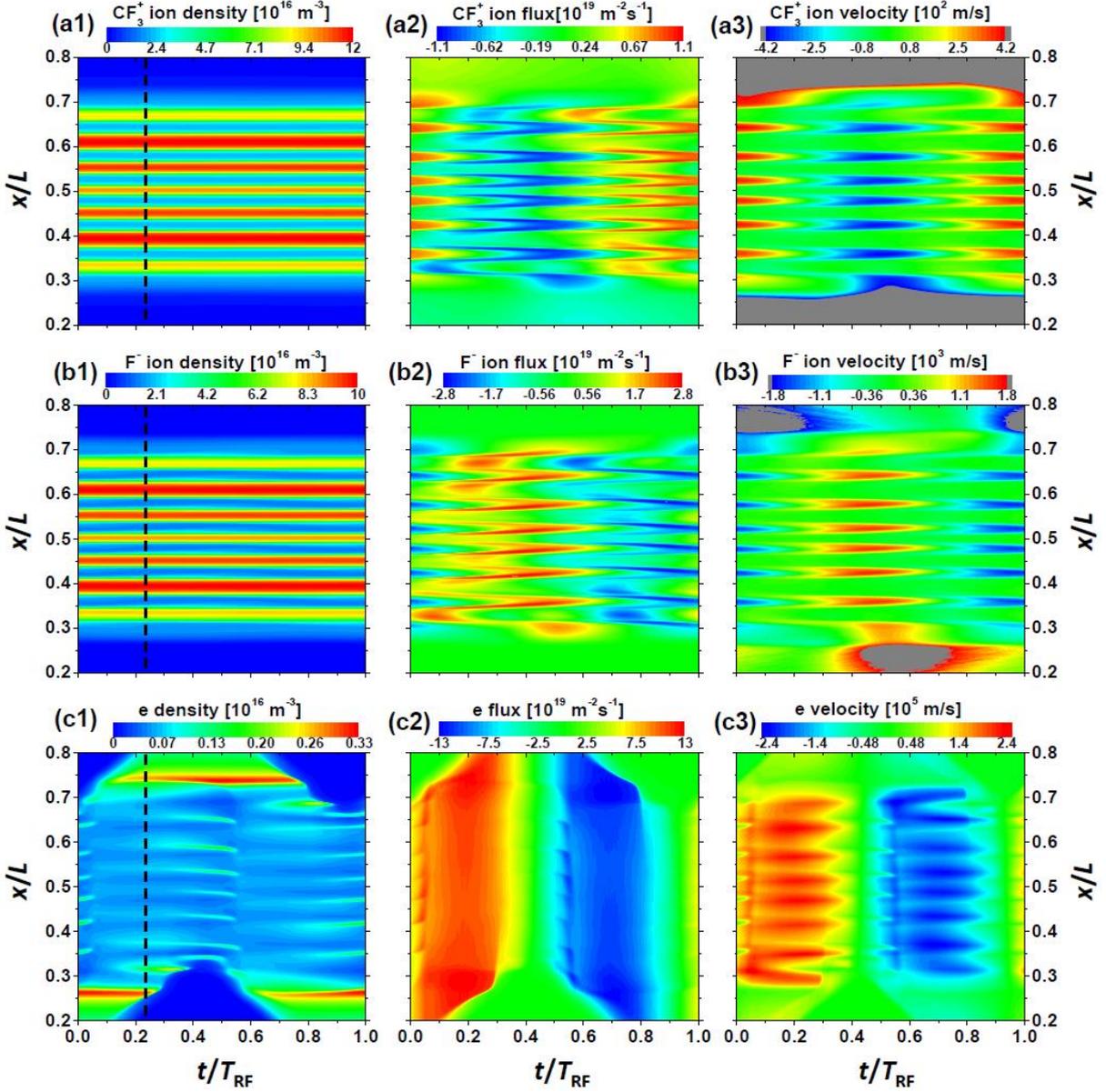

Figure 6 PIC/MCC simulation results: spatio-temporal plots of the $CF_3^+$ ion density, flux, and mean axial velocity (first row), $F^-$ ion density, flux, and mean axial velocity (second row), and electron density, flux, and mean axial velocity (third row) for the spatial region $0.2 \leq x/L \leq 0.8$ under the same conditions as in figure 4. Note that in the first column the vertical dashed lines indicate the time, at which the electron, $CF_3^+$, and $F^-$ ion density profiles in figure 5(a) are shown.

The spatio-temporal distributions of the electron density, flux, and mean velocity are shown in figures 6 (c1, c2, and c3), respectively. It can be seen that the electron density is strongly modulated in space. This spatial modulation develops after $t/T_{RF} = 0.3$, and maxima of the electron density form at the positions of $E = 0$ and where the electric field gradient is positive. In the bulk region the spatio-temporal profile of the electric field determines the temporal evolution of the electron density profile. The spatially modulated electron density profile, in turn, affects the local plasma conductivity, and consequently the electric field.

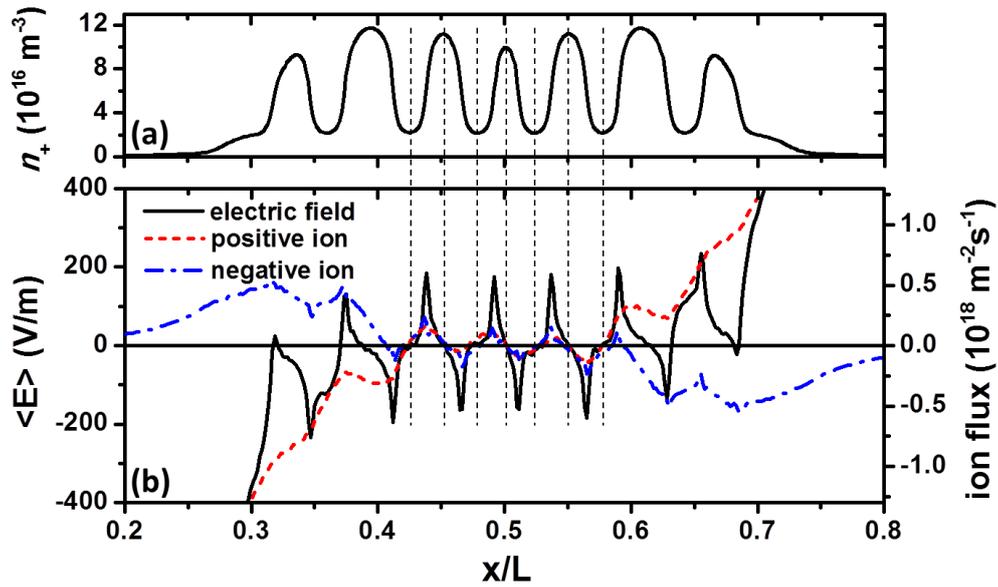

Figure 7 PIC/MCC simulation results: axial profiles of the time-averaged $CF_3^+$ ion density (a), time-averaged electric field $\langle E \rangle$ (black solid line), $CF_3^+$ ion flux (red dashed line) and $F^-$ ion flux (blue dashed-dot line) (b) for the spatial region $0.2 \leq x/L \leq 0.8$ under the same conditions as in figure 4.

The axial profiles of the time-averaged $CF_3^+$ ion density, time-averaged electric field $\langle E \rangle$, $CF_3^+$ and $F^-$ ion fluxes are shown in figure 7. In the bulk the time-averaged positive and negative ion fluxes are slightly modulated and both the positive and negative ions flow towards the ion density maxima on RF period average. The spatially modulated ion fluxes are caused by the spatial modulation of $\langle E \rangle$ in the bulk. The alternating regions of positive and negative electric field ensure that the local maxima of the ion densities (or comb-like ion density profile) remain stable, since these "focus" the positive ions to these positions of $E = 0$ with a negative field gradient and the more mobile negative ions follow the positive ion motion to maintain quasi-neutrality, since the positive ions are the majority species ($n^+ > n^-$).

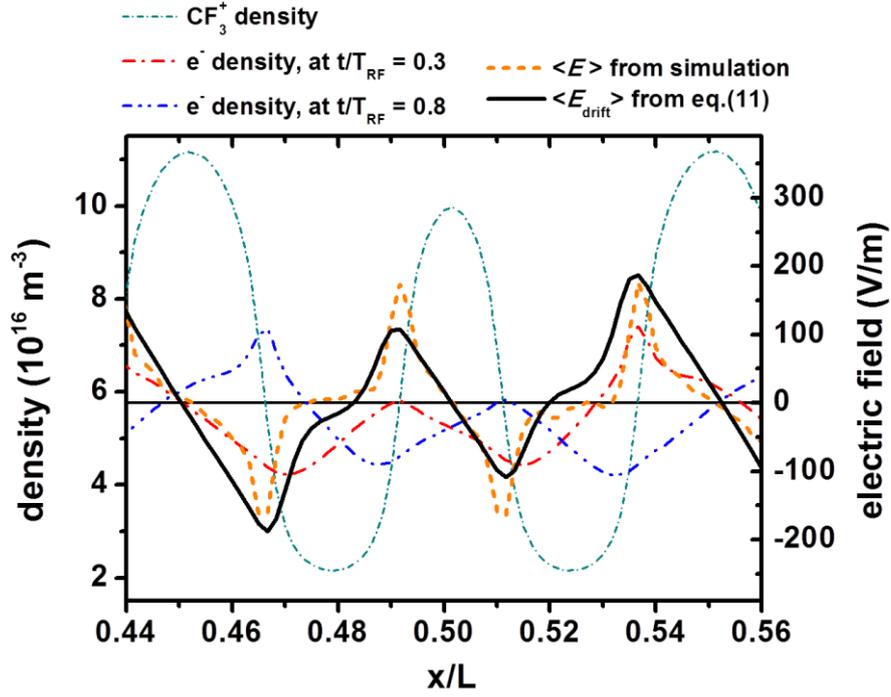

Figure 8 The axial profiles of the time-averaged $CF_3^+$ ion densities (teal dashed line), as well as the time-averaged electric field obtained from the simulation (yellow dashed line) and from the analytical model (black solid line, eq. (11)), respectively, and the axial profiles of the electron density at the times $t/T_{RF} = 0.3$ (red line) and $t/T_{RF} = 0.8$ (blue line) for the spatial region $0.45 \leq x/L \leq 0.55$ under the same conditions as in figure 4.

The underlying physics of the spatially modulated $\langle E \rangle$ is understood as follows. During the first half of one RF period, the transient electric field is negative between two ion density maxima, and is zero or slightly positive around the ion density maximum (see figure 5(a)). Upon responding rapidly to the bulk electric field, the electron density peaks at the positions, where the transient electric field vanishes, since the electrons are accelerated towards these positions. This leads to a locally-enhanced plasma conductivity ($\sigma \propto n_e$) and, consequently, to a momentary depletion of the absolute value of the local electric field at these positions, i.e., the electric field gets less negative there. This argument is based on the relationship $j_e \approx \sigma_{dc} \cdot E_{drift}$, where, $\sigma_{dc} = \frac{e^2 n_e}{m_e \nu_c}$ is the direct-current conductivity. During the second half of the RF period, the maxima of the electron density disappear at these positions, whereas some other peaks emerge at the left edges of the regions of the strong electric field (see blue dashed line in figure 8), since the electrons are accelerated towards these positions. Similarly, at these locations the plasma conductivity will be locally enhanced, and thus, the local electric field will become weaker (note that the electric field is positive for the second half of the RF period). On time average this mechanism leads to a spatially modulated time-averaged electric field in the bulk. To prove this mechanism the spatio-temporal distribution of $E_{\text{drift}}$ is estimated based on the relationship

$$E_{\text{drift}} \approx j_e \cdot m_e \nu_c / e^2 n_e. \tag{11}$$

where $j_e$ and $n_e$, are taken from the simulation, and the mean electron energy, $\langle \varepsilon_e \rangle$, is assumed to be $\langle \varepsilon_e \rangle = 5$ eV and used for the calculation of the collision frequency $\nu_c$. The time-averaged drift field $\langle E_{\text{drift}} \rangle$, obtained in this way is plotted in figure 8, and shows a general agreement with the time-averaged electric field $\langle E \rangle$ resulting directly from the simulation, suggesting that the electron dynamics is responsible for the spatially alternating $\langle E \rangle$, which plays an important role for the sustainment of the striations.

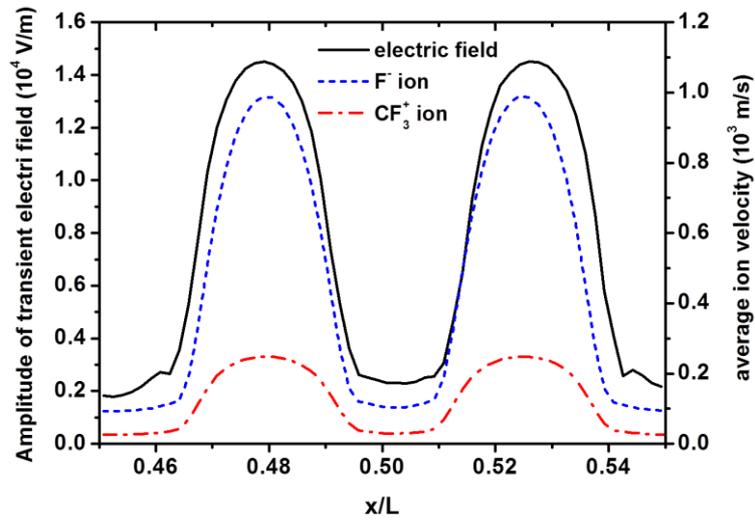

Figure 9 Axial profiles of the amplitudes of the transient electric field oscillation and the average velocities of $F^-$ and $CF_3^+$ ions for the spatial region $0.45 \leq x/L \leq 0.55$ under the same conditions as in figure 4. Note that the average ion velocities are obtained by time averaging the absolute value of the mean axial ion velocities shown in figures 6 (a3, b3).

In addition, the transient electric field is another important reason for the sustainment of the stable striations. Figure 9 shows the spatial profiles of the amplitudes of the transient electric field oscillation, as well as the average velocities of $F^-$ and $CF_3^+$ ions for the spatial region $0.45 \leq x/L \leq 0.55$. We find that within the region between two ion density maxima, the amplitude of the electric field oscillation exhibits a symmetrical parabola-like form, while at the locations of the ion density maxima the electric field is relatively flat and is much weaker. The amplitudes of the average ion velocities resemble that of the electric field, and the ion energies at the positions of the ion density maxima are typically low compared to these at the positions of ion density minima due to the relatively weak electric field there.

In the following, we consider the motion of a single ion in the spatially modulated electric field to analyze the role of the transient electric field for the sustainment of the striations. Ions residing originally at positions where the electric field is low and the ion density is high (at the positions of the striations), will experience a time-dependent electric field that confines them at the same positions as the amplitude of the transient electric field is increasing as a function of distance from the striations. For the same reason, an ion that is originally located between two striations, will be accelerated toward one of these, depending on its initial velocity. Therefore, besides the time-averaged field shown in figure 7, the parabola-like profile of the amplitude of the transient electric field is another reason for the sustainment of the striations. It is worth mentioning that the transient electric field can also play a key role in the development of the striations, since the spatially modulated transient electric field reinforces the response of the positive and negative ions to the RF electric field by pushing the positive and negative ions into the locations of the ion density maxima (as analyzed above). Consequently, the space charges, as well as the spatially modulated electric field are further enhanced due to the positive feedback. The effect is self-amplified until a periodic steady state is established.

Due to the fact that the ions are focused onto the closest striation, any ion losses in the volume (due to recombination) must be compensated by the generation of new ions within the adjacent striation gaps. This balance between particle generation and losses is expected to define the length of the striation gap, since this length affects the number of charged particle creation events in this region. For a given spatio-temporal distribution of the bulk electric field a larger striation gap will lead to an acceleration of electrons to higher energies and will increase their chances to create ions. The starting energy of the electrons is, in turn, affected by the striation width, which is determined by the ion displacement, since this width corresponds to the region of low electric field, where electrons lose energy by collisions (see figure 5). Thus, the maximum displacement of the ions is correlated with the striation gap: if the ion displacement increases, the striation gap will increase, since a larger region of high electric field in between two striation is required to generate enough ions to sustain the striation, because electrons enter the striation gap at lower energies (see section 3.2).

## 3.2 Effects of the driving frequency

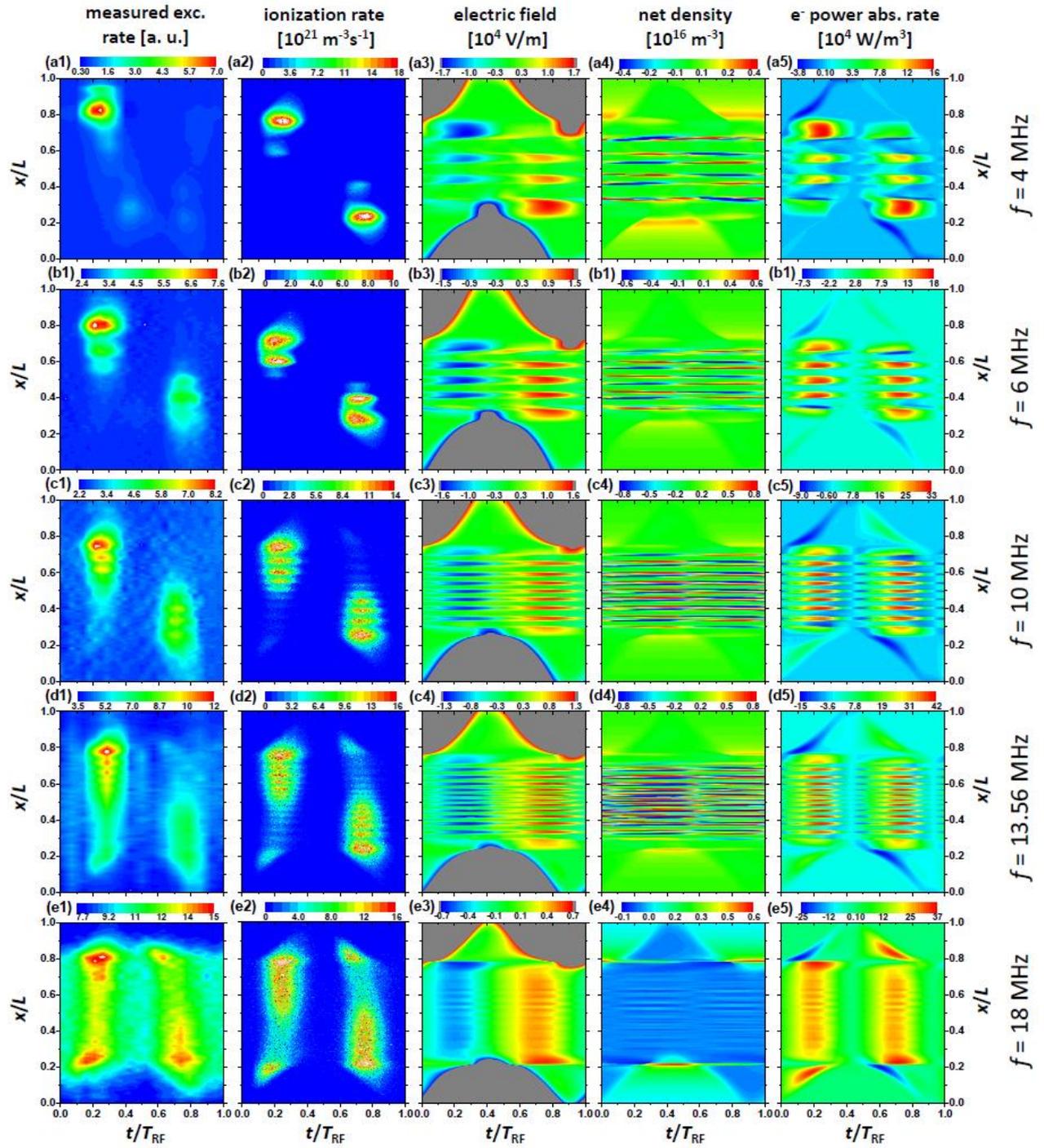

Figure 10. PROES results: electron-impact excitation rates at different driving frequencies (first column). PIC/MCC simulation results: the spatio-temporal plots of the electron-impact ionization rate (second column), electric field (third column), net charge density (fourth column) and electron power absorption rate (fifth column) at different driving frequencies. The other conditions are the same as in figure 4.

The effects of the driving frequency on the striated structure of different plasma parameters are investigated in this section. Figure 10 shows the driving frequency dependence of the spatio-temporal distributions of the measured electron-impact excitation rate and the calculated ionization rate, electric field, net charge density and electron power absorption rate under the same conditions (100 Pa, 300 V) as in figure 4. Generally, there is good agreement between the experimentally measured spatio-temporal distributions of the electron-impact excitation rates and the electron-impact ionization rates obtained from the simulation. As the driving frequency increases, the gap between the excitation/ionization rate maxima (defined as "striation gap") clearly decreases both in the experiment and in the simulation. At 4 MHz two excitation/ionization rate maxima can be observed during each half of one RF period. Compared to the simulation result at 4 MHz, in the experiment another excitation pattern occurs at the sheath edge, when the sheath is at its full expansion phase. This excitation pattern is caused by the secondary electron emission from the powered electrode due to the positive ion bombardment. The excitation pattern close to the powered electrode at the second half of one RF period is almost missing. This is because in the experiment a significant negative self-bias is generated due to the geometrical asymmetry of the plasma reactor. As the driving frequency increases to 6 MHz, one can observe three excitation/ionization rate maxima within each half RF period in the experiment/simulation and a greatly weakened excitation pattern caused by secondary electrons in the experiment, which vanishes at higher frequencies. At 13.56 MHz, the striation gap becomes quite narrow, and it cannot be spatially resolved at frequencies higher than 13.56 MHz in the experiment via PROES, while the striated ionization patterns are still observable up to the driving frequency $f = 18$ MHz in the simulation at the current conditions (i.e., at 100 Pa pressure and 300 V voltage amplitude). At $f \geq 18$ MHz, the striations in the spatio-temporal excitation/ionization rate vanish completely, which will be explained in detail later. Meanwhile, with the increase of the driving frequency, the discharge exhibits a mode transition from the striated/drift-ambipolar mode into a combination of drift-ambipolar and α modes. This transition occurs, because as the driving frequency increases, the sheaths expand faster, leading to an enhanced electron power absorption rate, as well as a rapid increase of the electron density and consequently a decrease of the electronegativity.

At 18 MHz the striated structure of different plasma parameters almost disappears, because the maximum of the ion density inside the bulk region is close to the "critical density" so that the eigenfrequency of the ion-ion plasma is close to the driving frequency, $f = 18$ MHz (see below), i.e., the ions can no longer respond to the RF electric field, and, consequently, no more alternating space charges are formed.

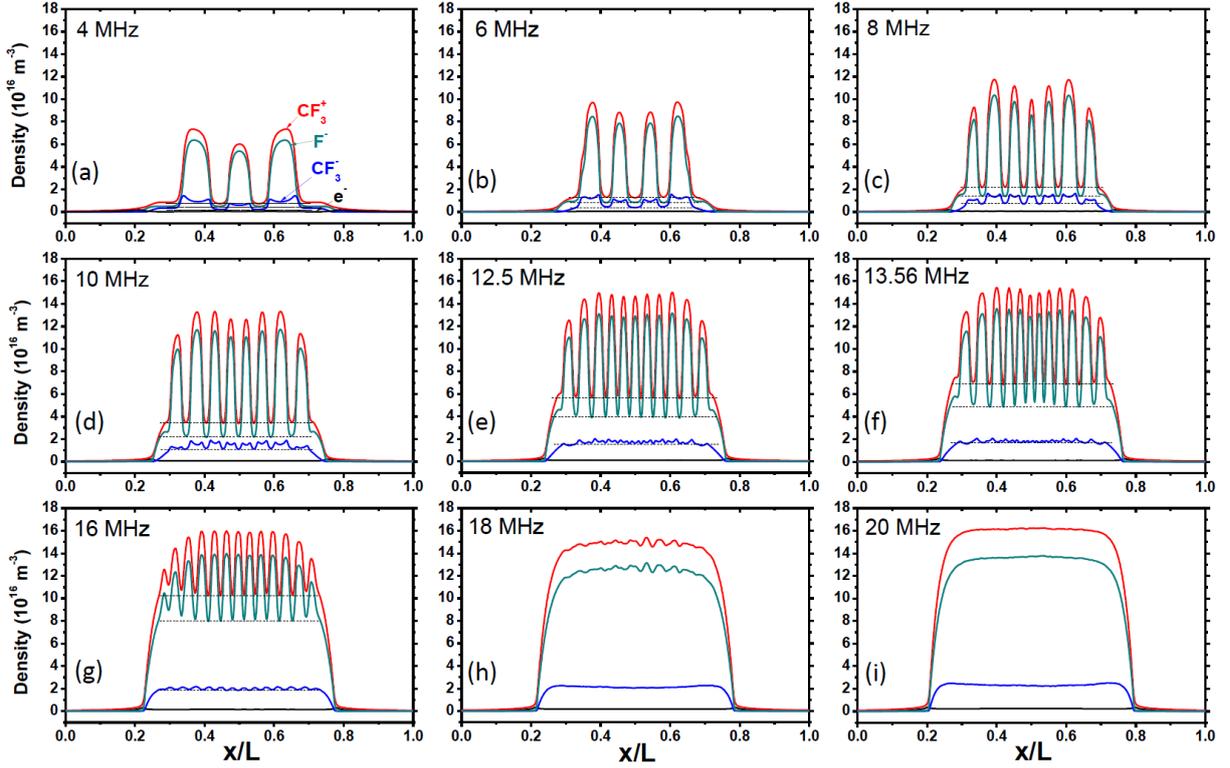

Figure 11. PIC/MCC simulation results: axial profiles of the time-averaged $CF_3^+$, $F^-$, $CF_3^-$ ion and electron densities in the driving frequency range of 4 ~ 20 MHz. The other conditions are the same as in figure 4.

Figure 11 shows simulation results for the axial profiles of the time-averaged charged species densities in the driving frequency range of 4 ~ 20 MHz. At $f < 18$ MHz, the $CF_3^+$ and $F^-$ ion densities exhibit comb-like profiles. Particularly, as the driving frequency increases the striation gap (i.e., the distance between two ion density maxima) decreases. At $f \geq 18$ MHz, the comb-like profile of the ion density disappears.

For a given value of the driving frequency when the striations are present, the minima of the densities of the major ionic species ($CF_3^+$ and $F^-$) are comparable in the bulk, as the minimum densities of the different ions are determined by the requirement to fulfill the resonance condition, which leads to minimum densities that are similar, but slightly different due to different degree in the spatial modulation of the electric field in the bulk plasma.

The maxima and minima of the $CF_3^+$, $F^-$ ion densities and the space-averaged electron density over the bulk region as a function of the driving frequency are presented in figure 12, which shows an abnormally enhanced maximum of the ion densities within the driving frequency range $4 < f < 18$ MHz that results from the "focusing effect".

In addition, by comparing the variations of the ion density minima and the calculated critical ion density $n_{critical}$ with the driving frequency in figure 12, we find that at $f < 18$ MHz the minima of the ion density change with the driving frequency in a similar manner as that of the critical ion density $n_{critical}$, which is calculated based on the resonance relationship $\omega_{rf}^2 = \omega^2 = e^2 n_{critical}/\varepsilon_0 \mu$. This indicates that at $f < 18$ MHz the ion density minimum exhibits an approximately quadratic increase with the driving frequency. For the same frequency domain the maximum ion densities are significantly higher than the critical density, i.e., striations can form. At $f > 18$ MHz, however, the maximum and minimum ion densities become comparable to or less than the critical ion density, i.e. striations cannot occur.

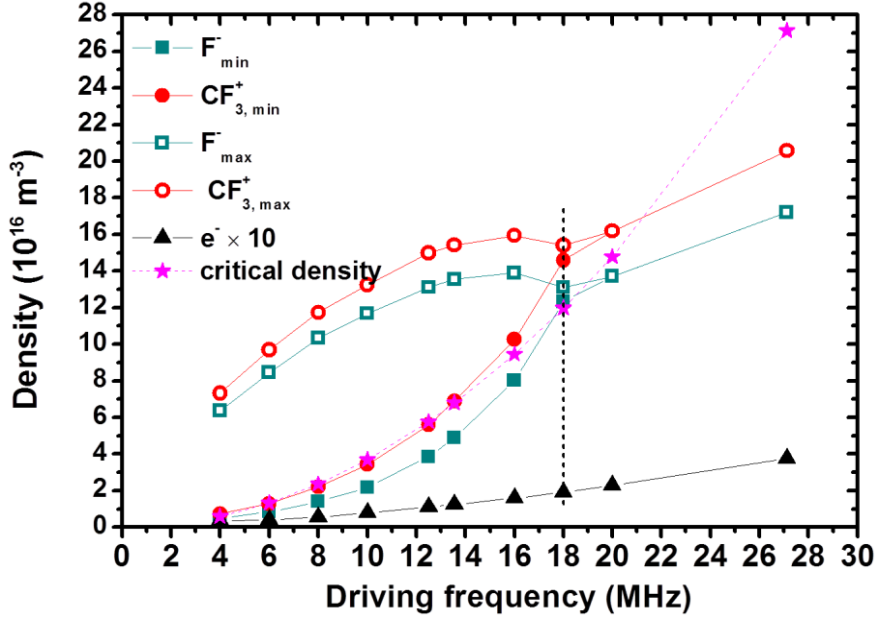

Figure 12. PIC/MCC simulation results: (a) the maximum and minimum of the $CF_3^+$ (red circles) and $F^-$ (teal rectangles) ion densities and the space-averaged electron density (black triangles) inside the bulk region (multiplied by a factor of 10) as a function of the driving frequency. The calculated critical ion density (pink stars) is also shown for comparison. The discharge conditions are the same as in figure 4.

Figure 13 shows the dependence of the local electronegativities at the positions of the maximum and minimum ion density (i.e., $\xi_{max}$ and $\xi_{min}$) on the driving frequency, $f$. At $f < 18$ MHz, when the striations are present, $\xi_{max}$ and $\xi_{min}$ exhibit opposite changes with $f$, i.e., $\xi_{max}$ generally decreases, while $\xi_{min}$ monotonously increases. This is because as $f$ increases, the electron density increases more rapidly than the maximum ion density in the bulk, and slower than the minimum ion density, which shows a quadratic increase with $f$. At $f > 18$ MHz, $\xi$ decreases as a function of $f$, suggesting that the discharge becomes less electronegative.

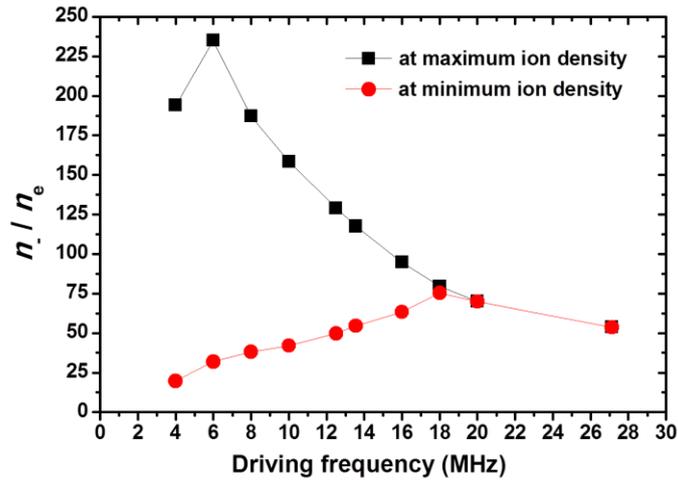

Figure 13 PIC/MCC simulation results: the local electronegativity, $\xi = n_-/n_e$, at the positions of the maximum (black rectangles) and minimum (red circles) ion density as a function of the driving frequency. The discharge conditions are the same as in figure 4.

Figure 14 shows the amplitude of the transient electric field oscillation (refer to figure 9), as well as the local maximum of the time-averaged field $\langle E \rangle$ at the locations close to $x/L = 0.5$ as a function of $f_{\text{driv.}}$. As it can be seen from figure 14, the amplitude of the transient electric field is remarkably enhanced in the frequency range between 4 MHz and 18 MHz, with a peak occurring around 10 MHz, indicating the most significant spatial modulation of plasma parameters. The local maximum of $\langle E \rangle$ changes with the driving frequency in a similar manner as the amplitude of the transient electric field does. As mentioned above, such a transient, or time-averaged electric field can focus the positive and negative ions into the locations of the local density peaks, leading to the reduction of the minimum ion density. At $f > 18$ MHz, the spatial modulation of the bulk electric field vanishes, and, as a consequence, the bulk electric field reduces to the spatially-uniform drift electric field and, thus, $\langle E \rangle$ tends to be zero (see figure 14).

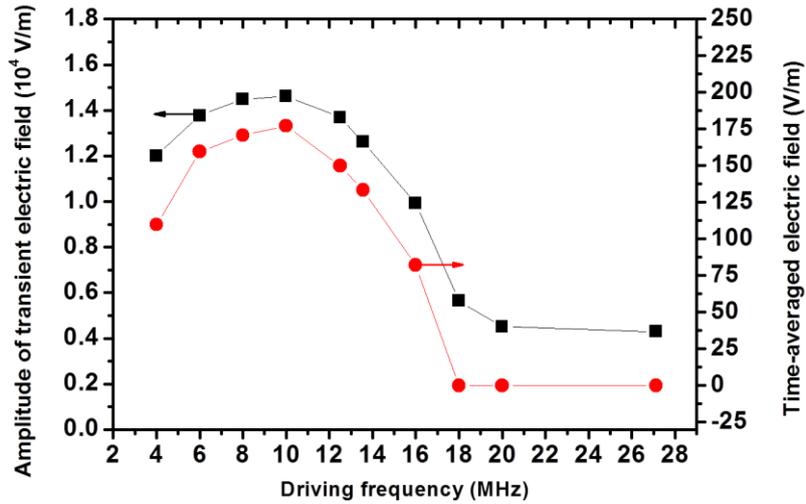

Figure 14. PIC/MCC simulation results: the amplitude of the electric field oscillation at the location of the central ion density minimum (black rectangles) and the average of the absolute values of maximum and minimum of the time-averaged electric field at the edges of the central ion density maximum (red circles) as a function of the driving frequency. The conditions are the same as in figure 4.

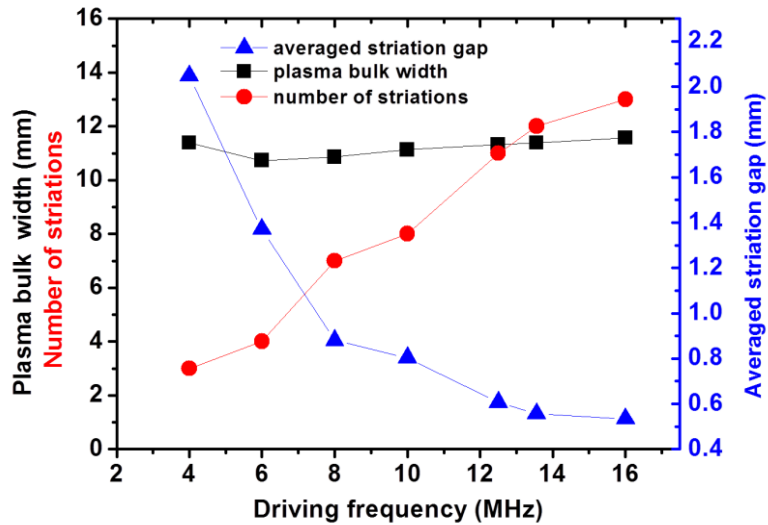

Figure 15 PIC simulation results: plasma bulk width (black rectangles), number of ion density peaks (red circles), and the average striation gap (blue triangles) as a function of the driving frequency. The discharge conditions are the same as in figure 4.

The dependence of the striation gap on the driving frequency is depicted in figure 15, in which the number of striations (see figure 12) and the plasma bulk width as a function of the driving frequency are also shown. Note here that the plasma bulk width $b$ satisfies $b = L - s_{\max}$, where $L$ is electrode gap, $s_{max}$ is the maximum of the sheath width. The sheath width is determined by Brinkmann's criterion [58]. We see that the bulk plasma width remains unaffected by the driving frequency, while the number of striations generally exhibits an approximately linear increase. So, it is apparent that the striation gap is approximately inversely proportional to the driving frequency, as shown in figure 15. We find that the striation gap is correlated with the maximum relative displacement of the $CF_3^+$ and $F^-$ ions per half RF period, i.e., the striation gap changes with the driving frequency in the same manner as the maximum relative displacement of positive and negative ions for each half RF period. The frequency dependence of the maximum relative displacement can be given as the amplitude of the analytical solution of the ion-ion plasma model, $A^* = \frac{eE_0/\mu}{\omega_{rf}\nu}$, according to which $A^*$ is inversely proportional to the driving frequency ($f_{rf} = \omega_{rf}/2\pi$). According to the discussion at the end of section 3.1 a larger ion displacement leads to broader ion density maxima, where the bulk electric field is reduced by the presence of the space charges formed at the edges of the ion density maxima. According to figure 5 electrons lose energy via collisions within these regions of low electric field. In order to sustain the striations the local charged particle losses must be balanced by the corresponding sources, e.g., electron impact ionization for the generation of electrons and positive ions, within the adjacent striation gaps, where the electric field is high. A larger ion displacement leads to a lower mean electron energy, where and when electrons enter a given striation gap. Thus, in order to generate sufficient ionization to compensate the losses the striation gap must increase and is correlated with the ion discplacement, which decreases as a function of the driving frequency according to the analytical model.

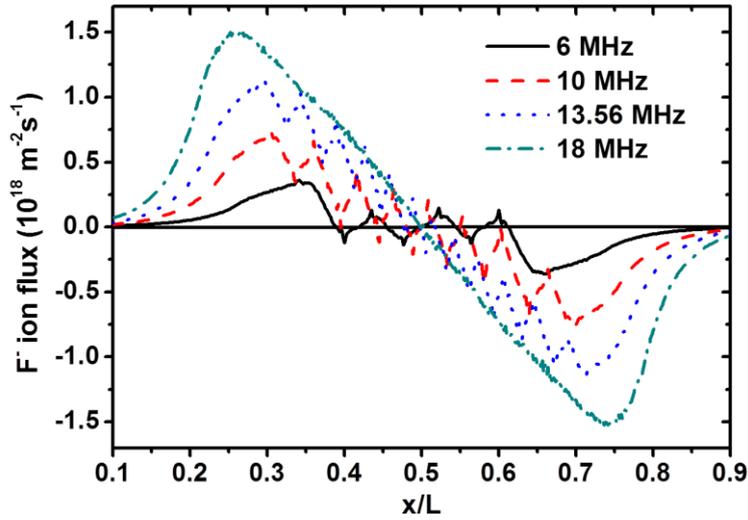

Figure 16. PIC/MCC simulation results: axial profiles of the time-averaged $F^-$ ion flux at different frequencies, 6, 10, 13.56, 18 MHz for the spatial region $0.1 \leq x/L \leq 0.9$ under the same conditions as in figure 4.

The axial profiles of the time-averaged $F^-$ ion flux for different driving frequencies and within the spatial region $0.1 < x/L < 0.9$ are shown in figure 16, from which one can deduce that the time-averaged $F^-$ ion flux generally exhibits similar axial profiles for different driving frequencies. However, as the driving frequency increases the peak of the $F^-$ ion flux at the edge of the bulk is enhanced, due to the increase in the $F^-$ ion density at a higher frequency. Besides that, we see that at $f < 18$ MHz there is a spatially modulated time-averaged $F^-$ ion flux in the bulk, which vanishes at $f = 18$ MHz (no striations). The missing spatial modulation of the time-averaged $F^-$ ion flux is due to the missing spatial modulation of the electric field inside the bulk region, when the striations are absent.

## 4. Conclusions

Self-organized striated structures of the plasma emission have previously been observed in capacitively coupled radio-frequency $CF_4$ plasmas by Phase Resolved Optical Emission Spectroscopy (PROES). Their formation and sustainment mechanisms have been understood by particle-based kinetic simulations. These striations have been found to be generated due to the resonance between the driving radio frequency and the eigenfrequency of the ion-ion plasma, which leads to the periodic formation of space charges and corresponding electric fields at distinct positions of ion density gradients in the plasma bulk. In this work, the spatially modulated charged species densities, net charge density, electric field, electron energy gain, and other plasma parameters at a base case (8 MHz driving frequency, 300 V voltage amplitude and 100 Pa pressure) have been discussed in detail. The effects of the driving frequency on the striated structure of various plasma parameters have been studied by PROES and via PIC/MCC simulations, as well as an ion-ion plasma model.

At the base case, the electron-impact excitation, ionization, dissociative attachment rates, the electric field, the net charge density, the electron power absorption rate, and the electron mean energy are strongly modulated in space. Compared to the electronic ionization patterns, which are concentrated within a narrow region close to the edge of the collapsing sheath, the electron impact excitation and electron attachment patterns spread over the entire bulk region due to their relatively low threshold energies. The maxima of the electronic ionization, excitation and dissociative attachment rates always occur at the edge of narrow regions of strong electric field inside the bulk region, because the electrons reach their maximum mean energy after traversing these regions. The spatial profile of the electron density is primarily determined by the electric field, however, it is so small in the plasma bulk that it hardly contributes to the total space charge profile. The time-averaged positive and negative ion fluxes are spatially modulated in the bulk, which ensures that the comb-like ion density profiles persist.

The experimentally determined electronic excitation patterns show an excellent qualitative agreement with the simulation results for different driving frequencies. The striations occur within a wide driving frequency range below 18 MHz. This is confirmed by the resonance condition predicted by the ion-ion plasma model. The striation gap is found to correlate with the decrease of the ion displacement as a function of the driving frequency predicted by the model. This correlation is explained by the effect of the striation width on the electron dynamics and the requirement to balance the generation and losses of ions locally within the region around a single striation. For a given driving frequency, the minima of the densities of the dominant ionic species ($CF_3^+$ and $F^-$) inside the bulk are similar to ensure the fulfillment of the resonance condition. These minima show an approximately quadratic increase with the driving frequency. For frequencies below 18 MHz the maximum ion densities are found to be above the critical density required for the formation of the striations and predicted by the model, while for frequencies above 18 MHz this is no longer the case. Therefore, striations form only for frequencies below 18 MHz. The striations are found to be maintained by a focusing effect of the ions into the striations by the time-averaged and transient electric fields.

This novel effect substantially changes the discharge structure of electronegative CCRF plasmas and could drastically affect various process relevant plasma parameters, e.g., the flux-energy distribution functions of different charged species. Therefore, it is expected that these structures can play an important role in various plasma-based applications, such as PECVD, usually performed at relatively high pressures, low driving frequencies, and in electronegative gases. The striated structure is also expected to be very sensitive to the type of ion species (i.e., the ion mass in electronegative gas plasmas such as $O_2$, $SF_6$, $C_4F_8$), collision frequencies (or working pressure), etc.

**Acknowledgments**

This work has been financially supported by the National Natural Science Foundation of China (NSFC) (Grants No. 11335004 and No. 11405018), by the Hungarian National Research, Development and Innovation Office, via Grant No. NKFIH 119357, by the German Research Foundation (DFG) within the frame of the collaborative research center SFBTR87, and by the US National Science Foundation (Grant No. 1601080).

**References**

[1] Kolobov V I 2006 Striations in rare gas plasmas, J. Phys. D **39** R487

[2] Radu I, R Bartnikas, Czeremuszkin G, and Wertheimer M R 2003 Diagnostics of dielectric barrier discharges in noble gases: Atmospheric pressure glow and Pseudoglow discharges and spatio-temporal patterns, IEEE Trans. Plasma Sci. **31** 411

[3] Nie Q Y, Ren C S, Wang D Z, Li S Z, and Zhang J L, Kong M G 2007 Self-organized pattern formation of an atmospheric pressure plasma jet in a dielectric barrier discharge configuration, Appl. Phys. Lett. **90** 221504

[4] Shirafuji T, Kitagawa T, Wakai T, and Tachibana K 2003 Observation of self-organized filaments in a dielectric barrier discharge of Ar gas, Appl. Phys. Lett. **83** 2309

[5] Muller I, Punset C, Ammelt E, Purwins H G, and Boeuf J P 1999 Self organized filaments in dielectric barrier glow discharges, IEEE Trans. Plasma Sci. **27** 20

[6] Yang Y, Shi J, Harry J E, Proctor J, Garner C P, and Kong M G 2005 Multilayer plasma patterns in

atmospheric pressure glow discharges, IEEE Trans. Plasma Sci., **33** 298

[7] Diebold D, Forest C E, Hershkowitz N, Hsieh M K, Intrator T, Kaufman D, Kim G H, Lee S G, and Menard J 1992 Double-layer-relevant laboratory results, IEEE Trans. Plasma Sci. **20** 601

[8] Liu Y X, Schüngel E, Korolov I, Donkó Z, Wang Y N, and Schulze J 2016 Experimental observation and computational analysis of striations in electronegative capacitively coupled radio-frequency plasmas, Phy. Rev. Lett. **116** 255002

[9] Kawamura E, Lieberman M A and Lichtenberg A J 2016 Standing striations due to ionization instability in atmospheric pressure He/$H_2$O radio frequency capacitive discharges, Plasma Sources Sci. Technol. **25** 054009

[10] Kaganovich I D, Fedotov M A, and Tsendin L D 1994 Ionization instability of a Townsend discharge, Sov. Phys.-Tech. Phys. **39**, 241

[11] Compton K T, Turner L A, and McCurdy W H 1924 Theory and experiments relating to the striated glow discharge in mercury vapor, Phys. Rev. **24** 597

[12] Goldstein R A, M A Huerta, and Nearing J C 1979 Stationary striations in an argon plasma as a bifurcation phenomenon, Phys. Fluids **22** 231

[13] Golubovskii Y B, Maiorov V A, Porokhova I A, and Behnke J 1999 On the non-local electron kinetics in spatially periodic striation-like fields, J. Phys. D **32** 1391

[14] Lu X, Naidis G V, Laroussi M, Ostrikov K 2014 Guided ionization waves: Theory and experiments, Phys. Reports **540** 123

[15] Fujiwara Y, Sakakita H, Yamada H, Yamagishi Y, Itagaki H, Kiyama S, Fujiwara M, Ikehara Y, and Kim J 2016 Observations of multiple stationary striation phenomena in an atmospheric pressure neon plasma jet, Japanese J. Appl. Phys. **55**, 010301

[16] Sigeneger F and Loffhagen D 2016 Fluid model of a single striated filament in an RF plasma jet at atmospheric pressure, Plasma Sources Sci. Technol. **25** 035020

[17] Stittsworth J A, and Wendt A E 1996 Striations in a radio frequency planar inductively coupled plasma, IEEE Trans. Plasma Sci. **24**, 125

[18] Takahashi K and Ando A 2014 Observation of stationary plasma striation and collimated plasma transport in a 100-kHz inductively coupled plasma discharge, IEEE Trans. Plasma Sci. **42**, 2784

[19] Denpoh K 2012 Particle-in-cell/Monte Carlo collision simulations of striations in inductively coupled plasmas, J. Appl. Phys. **51** 106202

[20] Iza F, Yang S S, Kim H C, and Lee J K 2005 The mechanism of striation formation in plasma display panels, J. Appl. Phys. **98** 043302

[21] Linson L M and Workman J B 1970 Formation of striations in ionospheric plasma clouds, J. Geophys. Res. **75** 3211

[22] Mottez F, Chanteur G, and Roux A 1992 Filamentation of plasma in the auroral region by an ion-ion

instability: A process for the formation of bidimensional potential structures, J. Geophys. Res. **97** 10801

[23] Penfold A S, Thornton J A, and Warder Jr R C 1973 Structured discharges in high frequency plasmas, Czech. J. Phys. B **23** 341

[24] Lieberman M A and Lichtenberg A J 2005 Principles of Plasma Discharges and Materials Processing 2nd edn (Hoboken, NJ: Wiley Interscience)

[25] Schulze J, Heil B G, Luggenhölscher D, Brinkmann R P, Czarnetzki U 2008 Stochastic heating in asymmetric capacitively coupled RF discharges, J. Phys. D **41** 195212

[26] Liu G H, Liu Y X, Wen D Q, and Wang Y N 2015 Heating mode transition in capacitively coupled $CF_4$ discharges: Comparison of experiments with simulations, Plasma Sources Sci. Technol. **24**, 034006

[27] Donkó Z, Schulze J, Czarnetzki U, Derzsi A, Hartmann P, Korolov I and Schüngel E 2012 Fundamental investigations of capacitive radio frequency plasmas: simulations and experiments, Plasma Phys. Control. Fusion **54** 124003

[28] Bruneau B, Gans Timo, O' Connell D, Greb A, Johnson E V, and Booth J P 2015 Strong Ionization Asymmetry in a Geometrically Symmetric Radio Frequency Capacitively Coupled Plasma Induced by Sawtooth Voltage Waveforms, Phy. Rev. Lett. **114** 125002

[29] Schulze J, Donkó Z, Schüngel E, Czarnetzki U 2011 Secondary electron in dual-frequency capacitively coupled radio-frequency discharges, Plasma Sourc. Sci. Technol. **20** 045007

[30] Schulze J, Derzsi A, Dittmann K, Hemke T, Meichsner J, and Donkó Z 2011 Ionization by drift and ambipolar electric fields in electronegative capacitive radio frequency plasmas, Phys. Rev. Lett. **107** 275001

[31] Lichtenberg A J, Kouznetsov I G, Lee Y T, Lieberman M A, Kaganovich I D, and Tsendin L D 1997 Modelling plasma discharges at high electronegativity, Plasma Sources Sci. Technol. **6** 437

[32] Midha V and Economou D 2001 Dynamics of ion-ion plasmas under radio frequency bias, J. Appl. Phys. **90**, 1102

[33] Kaganovich I D 2001 Negative ion density fronts, Phys. Plasmas **8** 2540

[34] Kuellig C, Wegner Th, and Meichsner J 2015 Instabilities in a capacitively coupled oxygen plasma Phys. Plasmas **22** 043515

[35] Teichmann T, Kuellig C, Dittmann K, Matyash K, Schneider R, and Meichsner J 2013 Particle-In-Cell simulation of laser photodetachment in capacitively coupled radio frequency oxygen discharges Phys. Plasmas **20** 113509

[36] Dittmann K, Kuellig C, Meichsner J 2012 Electron and negative ion dynamics in electronegative cc-rf plasmas, Plasma Phys. Control. Fusion **54** 124038

[37] Bruneau B, Lafleur T, Gans T, et al. 2016 Effect of gas properties on the dynamics of the electrical slope asymmetry effect in capacitive plasmas: comparison of Ar, $H_2$ and $CF_4$, Plasma Sources Sci. Technol. **25** 01LT02

[38] Derzsi A, Schüngel E, Donkó Z, Schulze J 2015 Electron heating modes and frequency coupling

effects in dual-frequency capacitive $CF_4$ plasmas, Open Chem. **13** 346

[39] Derzsi A, Donkó Z, Schulze J 2013 Coupling effects of driving frequencies on the electron heating in electronegative capacitive dual-frequency plasmas, J. Phys. D **46** 482001

[40] Schüngel E, Mohr S, Iwashita S, Schulze J and Czarnetzki U 2013 The effect of dust on electron heating and dc self-bias in hydrogen diluted silane discharges, J. Phys. D: Appl. Phys. 46 175205

[41] Tsendin L D 2010 Nonlocal electron kinetics in gas-discharge plasma, Phys-Usp. **53** 133

[42] Liu Y X, Zhang Q Z, Hou L J, Jiang W, Jiang X Z, Lu W Q and Wang Y N 2011 Collisionless bounce resonance heating in dual-frequency capacitively coupled plasmas, Phys. Rev. Lett. **107** 055002

[43] Liu Y X, Zhang Q Z, Liu J, Song Y H, Bogaerts A, and Wang Y N 2012 Effect of bulk electric field reversal on the bounce resonance heating in dual-frequency capacitively coupled electronegative plasmas, Appl. Phys. Lett. **101** 114101

[44] Schulze J, Donkó Z, Derzsi A, Korolov I, and Schüngel E 2015 The effect of ambipolar electric fields on the electron heating in capacitive RF plasmas, Plasma Sources Sci. Technol. **24** 015019

[45] Lee H. -C., Seo B. H., Kwon D. -C., Kim J. H., Seong D. J., Oh S. J., Chung C.-W., You K. H., and Shin C. H. 2017 Evolution of electron temperature in inductively coupled plasma, Appl. Phys. Lett. **110** 014106

[46] Gans T, O' Connell D, Schulz-von der Gathen V and Waskoenig J 2010 The challenge of revealing and tailoring the dynamics of radio-frequency plasmas, Plasma Sources Sci. Technol. **19** 034010

[47] J Schulze, E Schüngel, Z Donkó, D Luggenhölscher and U Czarnetzki 2010 Phase resolved optical emission spectroscopy: a non-intrusive diagnostic to study electron dynamics in capacitive radio frequency discharges, J. Phys. D: Appl. Phys. **43** 124016

[48] Matyash K, Schneider R, Taccogna F, Hatayama A, Longo S, Capitelli M, Tskhakaya D, and Bronold F X 2007 Particle in Cell Simulation of Low Temperature Laboratory Plasmas Contrib. Plasma Phys. **47**, 595

[49] Kurihara M, Petrovic Z Lj and Makabe T 2000 Transport coefficients and scattering cross-sections for plasma modeling in $CF_4$-Ar mixtures: a swarm analysis J. Phys. D: Appl. Phys. **33**, 2146

[50] Bonham R A 1994 Electron Impact Cross Section Data for Carbon Tetraflouride, Jpn. J. Appl. Phys. **33**, 4157

[51] Georgieva V, Bogaerts A and Gijbels R 2003 Numerical study of Ar/CF4/N2 discharges in single- and dual-frequency capacitively coupled plasma reactors, J. Appl. Phys. **94**, 3748

[52] Georgieva V, Bogaerts A and Gijbels R 2004 Numerical investigation of ion-energy-distribution functions in single and dual frequency capacitively coupled plasma reactors Phys. Rev. E **69**, 026406;

[53] Nanbu K 2000 Probability theory of electron-molecule, ion-molecule, molecule-molecule, and Coulomb collisions for particle modeling of materials processing plasmas and gases IEEE Trans. Plasma Sci. **28**, 971

[54] Proshina O V, Rakhimova T V, Rakhimov A T and Voloshin D G 2010 Two modes of capacitively coupled rf discharge in $CF_4$ Plasma Sources Sci. Technol. **19**, 065013

[55] Denpoh K and Nanbu K 2000 Self-Consistent Particle Simulation of Radio Frequency $CF_4$ Discharge: Effect of Gas Pressure, Jpn. J. Appl. Phys. **39**, 2804

[56] Kollath R 1956 Encyclopedia of Physics, Flügge S, Ed. vol. 21 p. 264 (Berlin: Springer)

[57] Brandt S et al. 2016 Electron power absorption dynamics in capacitive radio frequency discharges driven by tailored voltage waveforms in $CF_4$, Plasma Sources Sci. Technol. **25**, 045015

[58] Brinkmann R P 2007 Beyond the step model: approximate expressions for the field in the plasma boundary sheath, J. Appl. Phys. **102** 093302